%% file: main.tex
\documentclass[sigconf]{aamas} 
\usepackage{eso-pic}
\usepackage{totpages}

\usepackage{balance} 
\usepackage{subcaption} 
\usepackage{hyperref}

\hypersetup{
    colorlinks=true,  
    linkcolor=blue,   
    filecolor=magenta, 
    urlcolor=blue,    
}

\makeatletter
\gdef\@copyrightpermission{
  \begin{minipage}{0.2\columnwidth}
   \href{https://creativecommons.org/licenses/by/4.0/}{\includegraphics[width=0.90\textwidth]{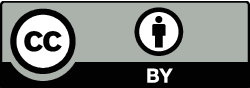}}
  \end{minipage}\hfill
  \begin{minipage}{0.8\columnwidth}
   \href{https://creativecommons.org/licenses/by/4.0/}{This work is licensed under a Creative Commons Attribution International 4.0 License.}
  \end{minipage}
  \vspace{5pt}
}
\makeatother

\setcopyright{ifaamas}
\acmConference[AAMAS '25]{Proc.\@ of the 24th International Conference
on Autonomous Agents and Multiagent Systems (AAMAS 2025)}{May 19 -- 23, 2025}
{Detroit, Michigan, USA}{Y.~Vorobeychik, S.~Das, A.~Nowé  (eds.)}
\copyrightyear{2025}
\acmYear{2025}
\acmDOI{}
\acmPrice{}
\acmISBN{}

\acmSubmissionID{<<992>>}

\title{Large Language Models for Virtual Human Gesture Selection}
\author{Parisa Ghanad Torshizi}
\affiliation{
  \institution{Northeastern University}
  \city{Boston}
  \country{United States}}
\email{ghanadtorshizi.p@northeastern.edu}

\author{Laura B. Hensel}
\affiliation{
  \institution{University of Glasgow}
  \city{Glasgow}
  \country{United Kingdom}}
\email{laura.hensel@glasgow.ac.uk}

\author{Ari Shapiro}
\affiliation{
  \institution{Flawless}
  \city{Los Angeles}
  \country{United States}}
\email{ariyshapiro@gmail.com}

\author{Stacy C. Marsella}
\affiliation{
  \institution{Northeastern University}
  \city{Boston}
  \country{United States}}
\email{s.marsella@northeastern.edu}

\begin{abstract}
    Co-speech gestures convey a wide variety of meanings and play an important role in face-to-face human interactions. These gestures have been shown to significantly influence the addressee’s engagement, recall, comprehension, and attitudes toward the speaker. Similarly, they have been shown to impact human and embodied virtual agent interaction. The process of selecting and animating meaningful gestures has thus become a key focus in designing embodied virtual agents. However, the automation of this gesture selection process poses a significant challenge. Prior gesture generation techniques have attempted to address this challenge in varied ways from fully automated, data-driven techniques -- which often struggle to produce contextually meaningful gestures -- to more manual approaches of crafting gesture expertise, which are time-consuming and lack generalizability. In this paper, we leverage the semantic capabilities of Large Language Models to realize a gesture selection approach that suggests meaningful, appropriate co-speech gestures. We first illustrate the information on gestures encoded into GPT4. Then we perform a study to specifically evaluate alternative prompting approaches for their ability to select meaningful, contextually relevant gestures and to align them appropriately to the co-speech utterance. Finally, we detail and demonstrate how this approach has been implemented within a virtual agent system, automating the selection and subsequent animation of the selected gestures for human-agent interactions.
\end{abstract}

\AddToShipoutPictureBG*{
  \AtPageLowerLeft{
    \rotatebox{90}{
      \raisebox{\dimexpr\paperheight-1cm\relax}{%
        \hspace{5cm}
        \textcolor{gray}{\Huge\textsf{This paper was submitted to AAMAS 2025 on October 16, 2024, and accepted.}}%
      }
    }
  }
}


\newcommand{\BibTeX}{\rm B\kern-.05em{\sc i\kern-.025em b}\kern-.08em\TeX}

\begin{document}




\maketitle
\section{Introduction} %
\input{introduction}
\section{Background}
\AddToShipoutPictureBG*{
  \AtPageLowerLeft{
    \rotatebox{90}{
      \raisebox{\dimexpr\paperheight-1cm\relax}{%
        \hspace{5cm}
        \textcolor{gray}{\Huge\textsf{This paper was submitted to AAMAS 2025 on October 16, 2024, and accepted.}}%
      }
    }
  }
}
\input{background}

\section{Approach}
\AddToShipoutPictureBG*{
  \AtPageLowerLeft{
    \rotatebox{90}{
      \raisebox{\dimexpr\paperheight-1cm\relax}{%
        \hspace{5cm}
        \textcolor{gray}{\Huge\textsf{This paper was submitted to AAMAS 2025 on October 16, 2024, and accepted.}}%
      }
    }
  }
}
\input{approach}

\section{Analysis} 
\AddToShipoutPictureBG*{
  \AtPageLowerLeft{
    \rotatebox{90}{
      \raisebox{\dimexpr\paperheight-1cm\relax}{%
        \hspace{5cm}
        \textcolor{gray}{\Huge\textsf{This paper was submitted to AAMAS 2025 on October 16, 2024, and accepted.}}%
      }
    }
  }
}
\input{results}
\section{Implementation}
\AddToShipoutPictureBG*{
  \AtPageLowerLeft{
    \rotatebox{90}{
      \raisebox{\dimexpr\paperheight-1cm\relax}{%
        \hspace{5cm}
        \textcolor{gray}{\Huge\textsf{This paper was submitted to AAMAS 2025 on October 16, 2024, and accepted.}}%
      }
    }
  }
}
\input{implement}
\section{Discussion}
\AddToShipoutPictureBG*{
  \AtPageLowerLeft{
    \rotatebox{90}{
      \raisebox{\dimexpr\paperheight-1cm\relax}{%
        \hspace{5cm}
        \textcolor{gray}{\Huge\textsf{This paper was submitted to AAMAS 2025 on October 16, 2024, and accepted.}}%
      }
    }
  }
}
\input{discussion}
\section{Conclusions and future work}
\AddToShipoutPictureBG*{
  \AtPageLowerLeft{
    \rotatebox{90}{
      \raisebox{\dimexpr\paperheight-1cm\relax}{%
        \hspace{5cm}
        \textcolor{gray}{\Huge\textsf{This paper was submitted to AAMAS 2025 on October 16, 2024, and accepted.}}%
      }
    }
  }
}

\input{conclusion}

\begin{acks}
This material is based upon work supported by NSF Grant \#2128743. Any opinions, findings, conclusions, or recommendations expressed are those of the authors and do not necessarily reflect the views of the NSF. 
\end{acks}


\balance
\bibliographystyle{ACM-Reference-Format} 
\balance
\bibliography{bib}
\AddToShipoutPictureBG*{
  \AtPageLowerLeft{
    \rotatebox{90}{
      \raisebox{\dimexpr\paperheight-1cm\relax}{%
        \hspace{5cm}
        \textcolor{gray}{\Huge\textsf{This paper was submitted to AAMAS 2025 on October 16, 2024, and accepted.}}%
      }
    }
  }
}


\end{document}

%% file: introduction.tex
People now regularly interact with embodied facsimiles of people. Graphics-based virtual humans and social robots with anthropomorphic features and behaviors engage users using the same verbal and non-verbal behaviors that people use when interacting with each other. 
These technologies exploit that the nonverbal behaviors of participants powerfully influence face-to-face interaction.  

In this paper, we focus on an integral part of such social interactions: co-speech gestures. Gestures convey a range of meanings and have a powerful impact on face-to-face interaction~\cite{kendon1997gesture,calbris2011elements,goldin2013gesture,mcneill1992hand}, impacting the speaker's persuasiveness as well as an addressee's comprehension, recall, engagement, and trust in the speaker ~\cite{tversky2009embodied,tversky2012time,bavelas1994gestures}. 

However, these impacts are dependent on the particular gestures being used and the context in which they occur~\cite[e.g.,][]{hostetter2011}. Clinicians, politicians, and comedians use different gestures because they seek to achieve different goals in different contexts~\cite{foley_nonverbal_2010, hall_hands_2016,seizer_uses_2011}. Beyond these differences in contexts and goals, there are also considerable cultural and individual differences in the use of gestures~\cite{ozer2020gesture,kita2020cross}. 

This richness makes gesture selection and animation a fundamental challenge in embodied agent research. To solve this challenge, researchers have focused on various automated approaches to select and generate virtual human gestures, often relying on data-driven or analysis-driven approaches~\cite[for an overview, see][]{saund2021gesture, nyatsanga2023}. 

With any approach, there are design factors that are critical to the realization of the virtual agent's gestures. Among these are requirements for the gestures to effectively convey the speaker's communicative intent. Further, we assume the virtual human should use gestures consistent with its role in an interaction.
Thus, a virtual human who is taking on the role of a labor union executive should not gesture like a comedian or clinician, nor should it use gestures based on some statistical average over differing roles and situations.

In this paper, we thus assume one is designing a virtual human to perform a specific role in some application. Leveraging recent work on exploring the ability of Large Language Models (LLMs) to predict gestures~\cite{hensel2023large}, we investigate the use of LLMs, in particular the standard GPT-4, in gesture selection with an additional focus on implementing this approach within a virtual human architecture. What makes LLMs particularly promising as a component in gesture selection is that they are trained on very large and highly varied corpora. As we demonstrate, that leads to LLMs encoding representations of both the linguistic properties critical to conveying meaning in gesture performances, as well as the relation of those phenomena to actual gestures. Furthermore, we explore and evaluate LLMs' ability to represent rheme and theme~\cite{mcneill1992hand,fries1984theme}, metaphor~\cite{grady1997foundations,cienki1998metaphoric}, image schemas~\cite{cienki2005image} and rhetorical structures~\cite{mann1987rhetorical}. 
GPT4's ability is assessed in the context of crafting a specific individual, a skilled presenter from a labor organization. The gesture selection approach is also implemented into an embodied virtual agent architecture allowing automation of gesture selection.

Our contributions include exploring GPT-4’s capabilities in selecting appropriate gestures, as well as using it to do discourse analysis to determine when to gesture. We also evaluated different prompting approaches. Furthermore, we automated the mapping from GPT-4's output into a behavior specification required by the character animation system to select the gesture animation. Finally, we implemented this within a virtual human framework.

%% file: background.tex
Research on gestures and their generation in virtual humans has a rich history, with gesture studies spanning millennia and virtual gesture generation predating the turn of the century (e.g.,~\cite{cassell1994animated}). In this section, we touch on some key aspects relevant to how we approached the problem of gesture generation.
Research in human co-speech gestures has identified different categorizations of gestures, related to the kinds of information they convey. For example, the work of McNeil (e.g.,~\cite{mcneill1992hand}) identifies four gesture categories—deictic, beat, metaphoric, and iconic.
This work will largely focus on metaphoric gestures for two primary reasons: First, our research often focuses on skilled presenters, and such professional speakers tend to use these gestures frequently. Second, metaphoric gestures can require deep analysis of the utterance to uncover relations to physical imagery, called image schemas, that underlie metaphoric gestures. Uncovering these relations presents a challenge for automated gesture selection.

Specifically, image schemas~\cite{Grady+2005+35+56} are recurring spatiotemporal relationships grounded in our body's interaction in the world that are argued to underlie common patterns of physical and abstract reasoning, as well as motivate linguistic metaphors and metaphoric gestures. Consider this example of the image schema of container from our data set: "with workers and employers putting a little bit of contribution from each into a trust fund". Note container is not only the collection, the "trust fund", but there is also a pattern of reasoning associated with container, putting things into the container. The speaker used a pattern of metaphoric gestures associated with container, a gesture denoting the container, along with gestures depicting placing things into it.

\textbf{Ideational Units:}
The above container example leads to an even more challenging issue. Gestures also occur in sequences where the individual gestures convey inter-related meanings. These gestural sequences are called ideational units~\cite{calbris2011elements} or gesture units~\cite{kendon1997gesture}.
This coupling plays important demarcative functions as well as helping to convey related meanings. These ideational units can be associated with image schemas as above but they can also establish relations between parts of an utterance, giving them a rhetorical structure.
Consider a communicative intent to convey a rhetorical contrast between two abstract concepts such as strongly different political views. 
Metaphorically, abstract concepts can be viewed as physical objects, with physical properties such as locations and physical distance conveying conceptual differences. Thus the rhetorical contrast between political views can be conveyed metaphorically by abstract deictic gestures that convey this separation by pointing to disparate regions in space. 
Such gestures pose difficult challenges for gesture animation because they set up the physical space of multiple gestures in a consistent fashion that relates physical space and motion to meaning as the above two examples illustrate.
\subsection{Gesture Selection}
The process of automating the selection of co-speech gestures for an utterance can be broken into two tasks. 
The first task is determining which segments of an utterance have co-speech gestures. Since gestures emphasize as well as convey meaning for parts of the utterance, identifying these segments is fundamental. We characterize this as {\em when to gesture}. Then there is the question of {\em what gesture to use}. In other words, what the gesture should convey, given the utterance and the context of the interaction. 

\textbf{When to gesture: } 
Gestures are tied to what the speaker seeks to emphasize~\cite{clough_role_2020}, which in turn transforms the meaning conveyed. 
Therefore, to determine when to gesture, it is critical for co-speech gesture generation to derive emphasis information about the utterance. A common approach is to use prosodic cues~\cite{guellai_prosody_2014, fares2023, marsella2013virtual}. However, this presumes that the spoken utterance that is driving gesture generation includes prosodic cues appropriate for the meaning the speaker seeks to convey. An alternative is a discourse analysis~\cite{mcneill2000growth} that can identify the {\em theme} or topic of a sentence and {\em rheme} or focus of the sentence, which is what is being said about the topic. The rheme provides new information about the topic and, critically, tends to be associated with gesture co-occurence~\cite{mcneill1992hand,cassell2001beat}. 

Gesture generation systems have used rheme analysis~\cite{cassell2000} as well as prosodic analysis~\cite{marsella2013virtual} to determine when to gesture. Assuming we choose to use rheme analysis to automate when to gesture, the question becomes how to quickly and efficiently do the analysis. In the context of the current work, we can specifically ask whether LLMs can do this kind of discourse analysis as part of a gesture selection process. 




\textbf{What gesture to use: }
There is a considerable variation in gesture types and usage frequency across individuals and contexts~\cite{kendon2004gesture, calbris2011elements}. 

As just one example of metaphoric gestures, consider the linguistic metaphor that abstract concepts can be physical objects~\cite{grady1997foundations}. All the physical properties and actions one takes on physical objects can be used to convey abstract meanings gesturally. Thus, concepts, as physical objects, can be gesturally grasped (understood), they can be thrown away (rejected), and they can be big (important). Two concepts, such as political orientation, as physical objects can have separate locations to contrast them or convey their differences. Sets of abstract concepts can have cardinality (large or small) and openness(closed or open sets). Set operations are apparent in gestural forms, adding elements, deleting, and union of sets.

This richness is not surprising given the richness of metaphors represented in our language~\cite{grady1997foundations} and argued to be integral to our thought processes~\cite{lakoff2008metaphors}. It also represents a central challenge to approaches to gesture generation. 



\section{Related Work}

As noted earlier, a key challenge in designing embodied conversational agents involves selecting appropriate gestures for the character, with various approaches proposed to address this challenge~\cite{neff2016hand}.

Rule-based methods~\cite{lee2006nonverbal,cassell2001beat} rely on predefined sets of rules that select from a predefined set of gesture animations. For instance, the Behavior Expression Animated Toolkit (BEAT) is a rule-based system that generates body movements and vocal intonations based on linguistic and contextual features of the text~\cite{cassell2001beat}. 

Mixed approaches~\cite{marsella2013virtual,lhommet2015cerebella, ravenet2018automating} integrate machine learning and ontologies to perform syntactic, semantic, and prosodic analysis of the utterance to infer communicative intent and then use handcrafted knowledge to map that intent to non-verbal behaviors such as head movements, gestures, and gaze. 

Although these rule-based and mixed systems offer designer control over gesture selection, they are limited to a fixed relationship between the properties of the speech they infer and gestures. Essentially, there is a trade-off between the designer's more explicit control over a character's use of gesture and both the burden placed on the developer or designer and the flexibility of the approach.

Data-driven approaches aim to learn a mapping between utterances and gestural motion from annotated corpora of gestural performances.  Work in this area relies on, for example, deep learning techniques. A recent review of this deep learning work~\cite{nyatsanga2023} has identified several key limitations, including a lack of designer control over the performance and the limited ability of current approaches to realize semantically meaningful gestures such as metaphoric gestures. 
Recent work has become more focused on the use of Transformer-based and diffusion-based generative models for the selection of gestures. DiffMotion~\cite{zhang2023diffmotion} generates gestures by integrating an LSTM with a diffusion model. DiM-Gesture~\cite{zhang2024dim} integrates a Mamba-based fuzzy feature extractor with Mamba- 2 diffusion architecture, to generate personalized full-body gestures.

Recent advancements have shifted towards Transformer-based and diffusion-based generative models for gesture selection. DiffMotion~\cite{zhang2023diffmotion} combines an LSTM with a diffusion model to generate gestures. Meanwhile, DiM-Gesture~\cite{zhang2024dim} utilizes a Mamba-based fuzzy feature extractor alongside a Mamba-2 diffusion architecture, enabling the creation of personalized full-body gestures.

These alternatives are solving different problems in some respects. Rule-based approaches continue to be used in applications, such as health applications, where there are significant ethical concerns and the designer seeks to retain control over the interaction, whereas machine learning-based approaches allow for more general, natural-looking behavior but with less control over what the nonverbal behavior conveys.
\subsection{Use of LLMs}
\AddToShipoutPictureBG*{
  \AtPageLowerLeft{
    \rotatebox{90}{
      \raisebox{\dimexpr\paperheight-1cm\relax}{%
        \hspace{5cm}
        \textcolor{gray}{\Huge\textsf{This paper was submitted to AAMAS 2025 on October 16, 2024, and accepted.}}%
      }
    }
  }
}
In this paper, we explore a different approach. Unlike an end-to-end machine learning approach that maps from utterance to animation, we explore the use of LLMs for gesture selection. We therefore go beyond traditional mixed and rule-based approaches in actually proposing gestures based on a more thorough analysis of the utterance by the LLM but stop short of end-to-end machine learning-based approaches that also generate the animation. This is both a limitation and a potential benefit in that it allows for carefully crafted gestures that effectively convey desired communicative intent. 

In addition, LLMs encode massive amounts of information across domains. Unlike the simpler word-based ontologies like Wordnet used previously in mixed approaches, or even the data used to train end-to-end approaches, the far more extensive data used to train LLMs allows them to encode contextual, semantic, and structural relations at the phrase, utterance and situational levels. LLMs also encode extensive amounts of information across scientific domains. GPT4, when asked, can define rheme and theme, metaphors, image schema, and metaphoric gestures.
As one example:

\begin{verbatim}
GPT4: Image schemas are recurring, dynamic patterns
of our perceptual interactions and motor programs ...
basic cognitive structures that arise from bodily
interactions with the world ... are understood to be
one of the fundamental building blocks of human thought
\end{verbatim}

\begin{verbatim}
GPT4: Container:|definition|, Path:|definition|, 
Source-Path-Goal:|definition|, Force:|definition|...
\end{verbatim}

LLMs also have representations of different gestural forms and styles used by speakers in different roles. They have for example knowledge of clinicians, their role, and their goals, and can even capture the variations in gestural styles among well-known individuals within these professions. Moreover, they can be prompted with examples that prime and integrate with their existing knowledge to improve their ability to select gestures. We asked GPT4 to report a repertoire of gestures that a clinician and a politician would use.

\begin{verbatim}
GPT4: Here's a repertoire of hand gestures that
clinicians often use:
Palms Open and Upward: Indicates openness and
transparency, used when explaining procedures or 
discussing options.
Hand to Chest: Expresses sincerity and empathy, 
particularly in sensitive conversations.
Enclosed Hands or Fingers Interlocked: Demonstrates
patience and focus, typically seen during attentive
listening.
\end{verbatim}

\begin{verbatim}
GPT4: Here's a repertoire of hand gestures that 
politicians often use:
Fist Pumping: Used to express determination or 
victory, energizing the audience.
Thumb and Index Finger Pinch: Known as the 
"precision grip," it emphasizes specific and
important points, suggesting control.
Hand Chopping: Aligns with making strong, 
decisive points, conveying clarity and decisiveness.
\end{verbatim}

Detailing the extent of all these representations is beyond the scope of the current paper. Rather, the focus of this paper is to lay out an integrated approach to using LLMs to automate gesture selection and evaluate the ability to determine {\em when to gesture} and {\em what gestures to use}. 
Our approach builds on the work of Hensel et al.~\cite{hensel2023large} but goes further in exploring different prompting techniques, having the LLM also determine when to gesture, develop a prompting approach that allows the automated mapping of LLM output to a behavior specification for a character animation system and finally implementing the approach. The implementation takes utterance text as input, generates the utterance audio, and drives a virtual human's spoken dialog and gesturing.

%% file: approach.tex
To explore a gesture generation framework using GPT-4, we lay out a set of research questions relating to gesture generation more broadly and the selection of individual gestures more specifically:

{\em RQ1: Selection of Gestures.} What are the impacts of alternative prompting approaches on the appropriateness of gesture selection and the speed of inference?
\begin{itemize}
    \item {\em Appropriateness} How appropriate are the selected gestures, with regards to the context of the speech and the speaker?
    \item {\em Speed of inference} Can GPT-4 select gestures with a speed sufficient to enable real-time inference in animation systems?
\end{itemize}

{\em RQ2: Selection of when to gesture.} Does the rheme identified through GPT-4-enabled rheme and theme analysis correspond accurately to the actual gestural timings of the speaker in the data?
%
To answer these research questions, we first annotated hand gestures in a specific speech. We then used these annotations as ground truths to evaluate gestures suggested by GPT-4 on the same utterances. We used GPT-4 as its performance is superior to that of GPT-3 or GPT-3.5, as shown by~\cite{hensel2023large}. Specifically, we used the GPT-4 Chat Completion API as it is faster and more versatile in terms of parameter settings, with the following parameters: (Temperature: 0.2, Max tokens: 256, Frequency penalty: 0). The transcribed speech was split into utterances, using  end of sentence punctuations.
\subsection{Data}
To derive a set of ground truth gestures and utterances, we selected a video featuring Elizabeth Shuler, a labor activist, speaking at the Working Families Summit, which is available through the National Archives and Records Administration.\footnote{https://youtu.be/-6NA1xl32uY?si=lOjTOkyIrQhgnsKJ} To collate Elizabeth Schuler's gesture repertoire, we annotated three segments of the video in which she was the primary speaker, resulting in a total of six minutes of annotated video. We then split these annotations into a training set (21 utterances) and a test set (20 utterances). 
\subsection{Classes of Gestures: Gestural Intents}
Based on the training data set, we created a list of gestural classes based on image schemas and metaphors that were apparent in the speaker's performance. 
Each of these classes conveys a particular intent and we thus refer to these as \emph{gestural intents}. Note that alternative gestures with different physical properties can realize one of these gestural intents. For example, the speaker may convey the gestural intent of progress either through a forward circling of the hands or a sweep showing the direction of progress. Below, we list the speaker's repertoire of {\em gestural intents}, including brief explanations.
\begin{itemize}
    \item Progress: This gesture represents progress, advancement, or moving forward. It is part of the path group image schema.
    \item Regress: This gesture represents moving backward, regressing, or returning to a previous point. It is part of the path group image schema.
    \item Cycle: This gesture represents actions or processes that repeat in a continuous loop or follow a recurring pattern. It is an image schema.
    \item Collect: This gesture represents gathering, collecting, or bringing things together, into one entity. It is an image schema.
    \item Container: This gesture represents a boundary, a sweep, or an imaginary box holding a collection of items. This is an image schema and basis for the container metaphoric gesture.
    \item Oscillation: This gesture represents alternation, uncertainty, indecision, or items being out of balance. It is part of the balance group image schema.
    \item Temporal: There are many, culture-specific time metaphors. Here we refer to representing time as a line, with different points on the line representing past, present, and future.
\end{itemize}
\subsection{Experimental Design: Gesture Selection}
To evaluate the selection of gestures, we designed a 2X2 factorial design of alternative prompting approaches: \emph{gestural intents repertoire explained/not explained} \textbf{X} \emph{annotated examples given/not given}. For each of those approaches, we asked GPT-4 to produce gestures based on the utterances in the test set. Below is a detailed description of each approach.\\
\emph{Approach 0:} In this approach, the model is not prompted with any prior information, on gestural intents or annotations. The model is just asked to report a gesture for each of the utterances, describe the physical properties of its suggested gesture, and the specific phrase in that utterance that the gesture is trying to illustrate. \\
\emph{Approach 1:} In this approach, the model is only prompted with the above-mentioned list of gestural intents, and their descriptions. The model is then asked to report the gestural intents in the utterance, suggest a gesture for each gestural intent, the physical properties of that gesture, and the associated phrase.\\
\emph{Approach 2:} In this approach, the model is prompted with only the annotations from the training set.  The model is then asked to report the gestural intent, suggest a gesture for that gestural intent, the physical properties of that gesture, and the associated phrase.\\
\emph{Approach 3:} In this approach, the model is prompted with the above-mentioned list of gestural intents, and the annotations from the training set. The model is then asked to report the gestural intent, suggest a gesture for that gestural intent, the physical properties of that gesture, and the associated phrase. 


\subsection{Rheme/Theme Analysis: When to Gesture} 
We used GPT-4 driven discourse analysis to identify the rheme and theme of each utterance. We then explored whether the rheme and theme can serve as criteria to decide on when to gesture, or which part of the utterance is most likely to be accompanied by a gesture. Specifically, we prompted GPT-4 to identify the rheme and theme in each utterance. Such a rheme and theme analysis could be used on top of the gesture selection system, to prioritize which gesture/s to use in the animation system; with gestures associated with the rheme part of the utterance being given higher priority. To evaluate, we analyzed whether they correspond to the part of the utterance that the speaker gestured on.

%% file: results.tex
This section details the results of our experiments on GPT-4's ability to select appropriate gestures and determine when to gesture.
\subsection{Selection of Gestures}
In this section, we investigated our RQ1, examining the appropriateness of gestures selected by alternative prompting approaches and comparing their speed of inference. 
\subsubsection{Appropriateness}
To assess the semantic appropriateness of the gestures selected by the model, two experts jointly evaluated each proposed gesture for every utterance in the test set across all four approaches. Evaluators made their judgments for each gesture and then immediately compared them. If there was disagreement, the two discussed where this disagreement arose and resolved it through further discussion. It should be noted, however, that there was very little disagreement between the two experts.
In order to do the evaluation, we divided all the selected gestures by GPT-4 into the following two categories.\\
\textbf{Category 1:}\\
This category consists of gestures that the model generates for specific parts of an utterance, where there is a corresponding gesture on that (or close) part of the utterance in the actual speech (i.e., ground truth is available). Experts have assigned one of the following tags to these gestures:
\begin{itemize}
    \item \textbf{Appropriate}: The gesture does convey the semantic meaning that the speaker is trying to convey. Regarding the definition of appropriateness, we defined this as gestures that are not only appropriate for the utterance (i.e., ‘fit’ with the intended meaning) but also for the context of the speaker.
    \item \textbf{Inappropriate}: The gesture does not convey the semantic meaning that the speaker is trying to convey.
    \item \textbf{No corresponding gesture}: The speaker produced a gesture but the model did not propose a gesture at the corresponding part of the utterance.
\end{itemize}

\begin{figure}[t]
\centering
\begin{subfigure}[b]{0.37\textwidth}
    \includegraphics[width=\textwidth]{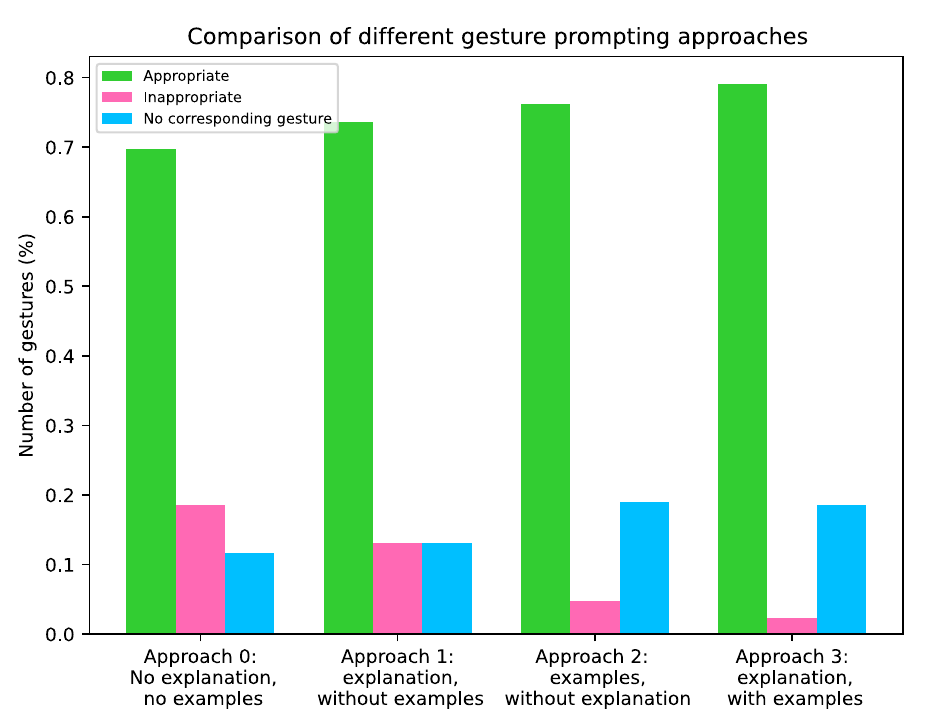}
    \Description{The appropriateness of gestures increases incrementally from approach 0 to approach 3. Also, the number of inappropriate gestures reduces incrementally from approach 0 to approach 3.}
    \caption{Selected gestures whose utterances have ground truth}
    \label{fig:sub1}
\end{subfigure}
\hfill 

\begin{subfigure}[b]{0.37\textwidth}
    \includegraphics[width=\textwidth]{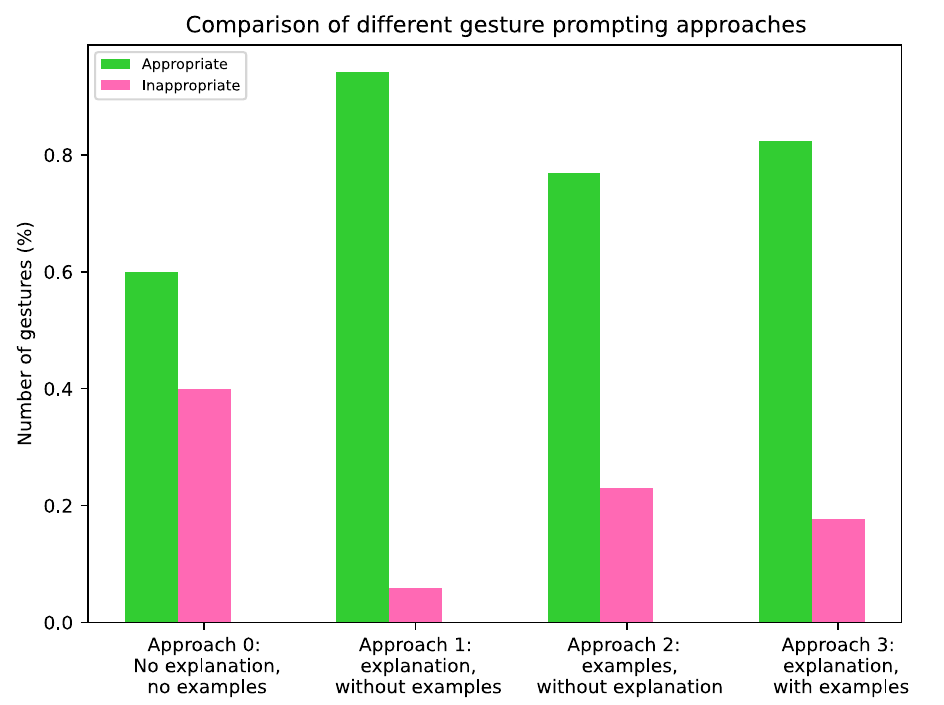}
    \Description{Approach 1 has the highest number of appropriate gestures, then approach 3, then approach 2, and finally approach 0 .In addition, approach 0 has the highest number of inappropriate gestures, then approach 2, then approach 3, and finally approach 1.}
    \caption{Selected gestures whose utterances do not have ground truth}
    \label{fig:sub2}
\end{subfigure}
\hfill 
\begin{subfigure}[b]{0.37\textwidth}
    \includegraphics[width=\textwidth]{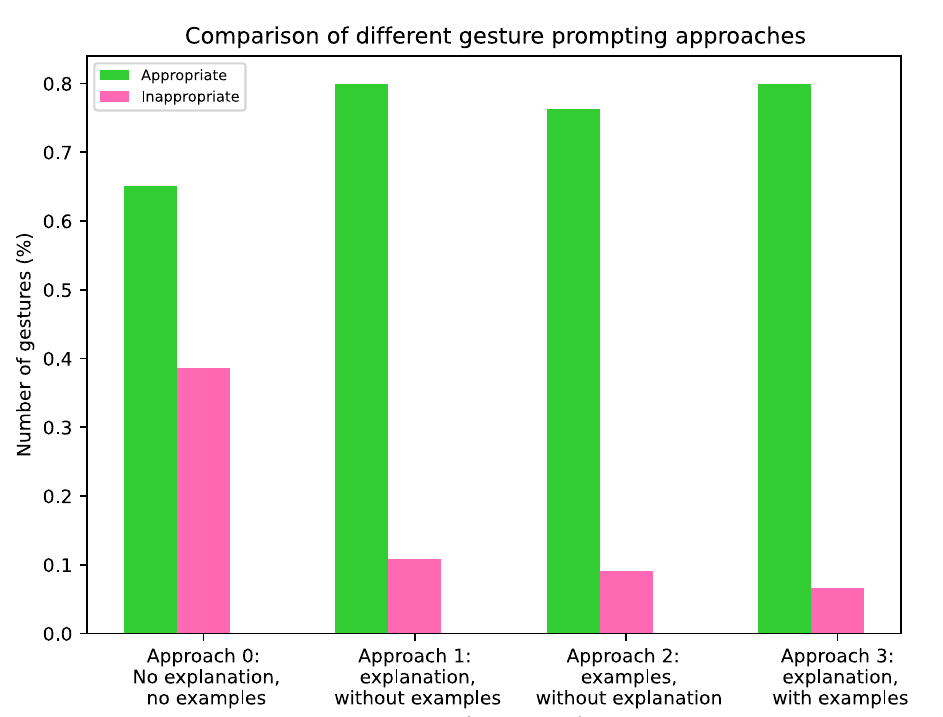}
    \Description{Approach 1 has the highest number of appropriate gestures, followed by approach 3,2, with a slight difference. Approach 0 has significantly higher inappropriate gestures.}
    \caption{all gestures}
    \label{fig:sub3}
\end{subfigure}
\caption{comparison of different prompting approaches in terms of their appropriateness}
\label{fig:subfigures}
\end{figure}

\begin{figure*}[ht]
\centering
\begin{subfigure}[b]{0.37\textwidth}
    \includegraphics[width=\textwidth]{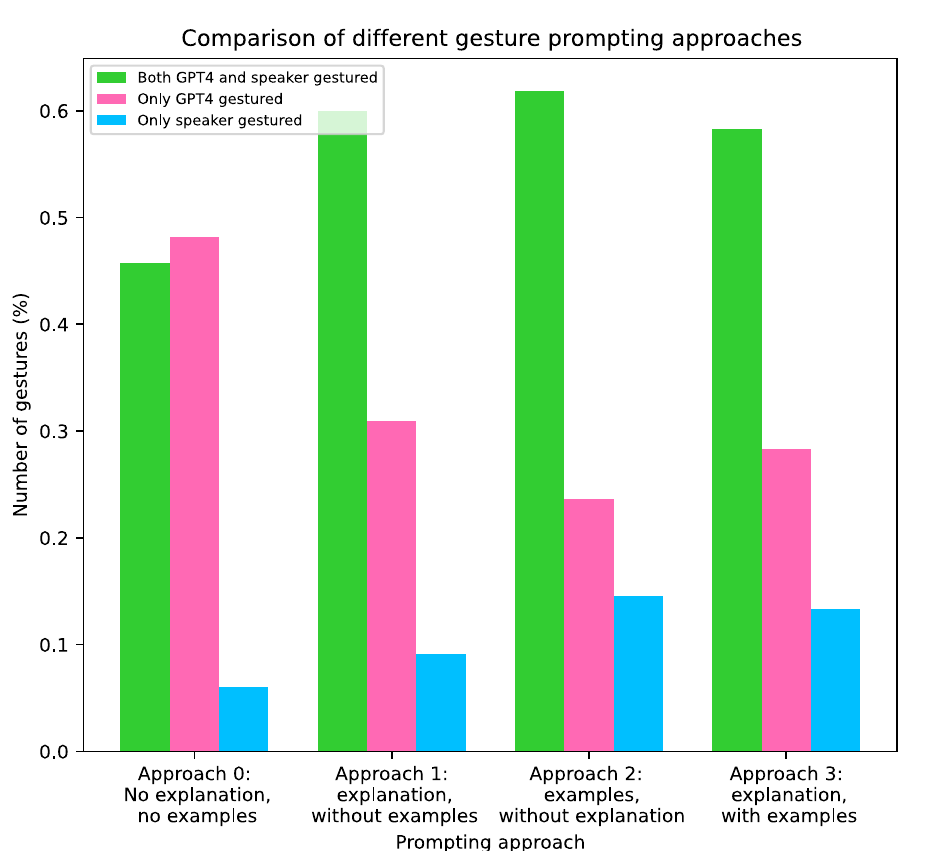}
    \Description{Approach 2 has the highest alignment between gestures suggested by the model and the speaker's gestures. Approach 0 has the lowest alignment.}
    \caption{Both speaker and GPT-4 gestured, GPT-4 gestured, \\ or speaker gestured}
    \label{fig:sub4}
\end{subfigure}
\begin{subfigure}[b]{0.37\textwidth}
    \includegraphics[width=\textwidth]{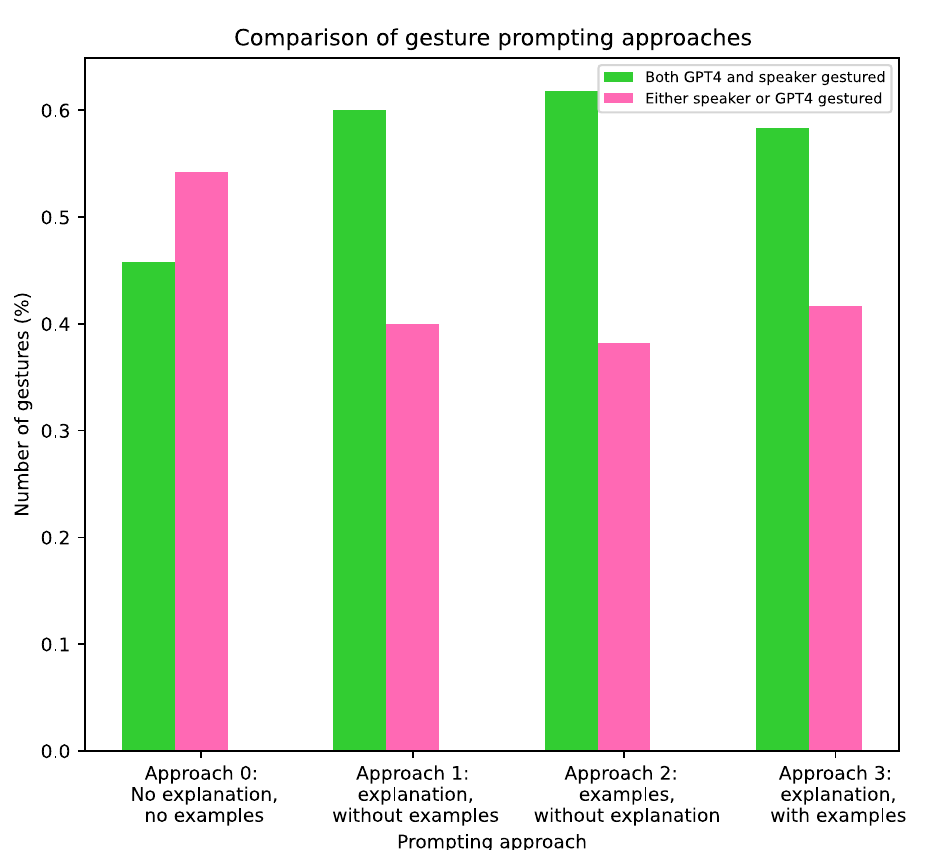}
        \Description{Approach 2 has the highest alignment between gestures suggested by the model and the speaker's gestures. Approach 0 has the lowest alignment.}
    \caption{Both speaker and GPT-4 gestured, either GPT-4 gestured or speaker gestured}
    \label{fig:sub5}
\end{subfigure}
\caption{comparison of different prompting approaches in terms of their alignment with the speaker}
\label{fig:subfigures2}
\end{figure*}

Figure \ref{fig:sub1} shows the number of gestures that belong to each label across the proposed prompting approaches. The results indicate that providing more information to the prompts yields more appropriate gestures and less inappropriate gestures. Specifically, approach 3 is better than approaches 1 and 2, and approaches 1,2, and 3 are better than approach 0. Moreover, providing the model with examples (approach 2) is slightly more effective in selecting more appropriate gestures compared to providing the model with solely the explanation of the gestures (approach 1).\\
\textbf{Category 2:}\\
This category consists of gestures that the model generates for specific parts of an utterance, yet there is no corresponding gesture on that (or close) part of the utterance in the actual speech (i.e., ground truth is not available). The experts labeled each gesture with either the appropriate or the inappropriate tag. \\
The results in Figure \ref{fig:sub2} suggest that approach 1 has the highest number of appropriate gestures and the lowest percentage of inappropriate gestures.\\
Combining all the gestures in the two categories listed above, we analyzed the overall appropriateness, regardless of whether a corresponding ground truth was present or not. The results, shown in Figure \ref{fig:sub3}, indicate that there is a slight difference between approaches 1,2 and 3, but they all outperform approach 0. Additionally, the performance of approach 1, where the model was not prompted with any annotations, is promising. This suggests that LLMs can minimize the need for extensive annotations.\\
Furthermore, we analyzed how often the gestures generated by GPT-4, regardless of their appropriateness, were aligned with the gestures performed by the speaker, meaning how often the speaker and GPT-4 gesture on the same phrase. Figure \ref{fig:sub4} and \ref{fig:sub5} depicts the frequency of gestures being aligned, or not aligned. In not-aligned cases, GPT-4 either produced gestures where the speaker did not gesture or the speaker gestured but GPT-4 did not produce a gesture. Approach 2 has a higher alignment of gestures between GPT-4 and the speaker. On the other hand, approach 0 produced the highest misalignment. These results suggest that limiting the gesture classes to the speaker's gesture repertoire helps to constrain gesture selection to produce more accurate gesture timings.
In conclusion, the approaches that prompted the model with a defined set of gestural intents generated more appropriate gestures and less inappropriate gestures. Moreover, they helped in selecting gestures that were happening concurrently with the speaker's gestures and thus were more aligned.

\subsubsection{Inference time}
As mentioned earlier, a potential use case for LLMs in gesture selection is to suggest gestures in real-time. Therefore, we explored the inference times of gesture selection across different prompting approaches. In this work, the inference times are actually recorded as network latencies, to evaluate real-time suitability for human-agent interactions. In these approaches, where GPT4 was queried to report gesture name, its properties, gesture description, and the associated phrase; the average inference time per utterance across different approaches ranges from 6.51 to 9.29 seconds. 
Specifically, approach 0 showed the highest inference time, and approach 1 had the lowest inference time. However, for the purpose of animating gestures in the virtual agent, since the animations are already defined, we truncated the query into asking only the gesture type and the associated phrase. We calculated different inference times for different models, GPT4 1.20 (s), GPT4-O-mini 1.08(s), and  Llama 3.1 1.23(s), Llama 2.7b fine-tuned on our dataset 1.8(s).
With a target inference time of less than one second for online, real-time, interactions, our next step is to adapt a smaller, faster model that can achieve this speed.

%
\subsection{Selection of When to Gesture: Rheme and Theme}
Next, in order to answer our RQ2, we sought to analyze how often the speaker's gestures occur within the rhemes identified by GPT-4. To do so, we considered all the gestures made by the speaker, including deictics, beats, metaphorics, and iconics; and counted the number of times these gestures happened within the rheme of the utterances identified by GPT-4. \\
The results show that out of 49 gestures made by the speaker, 43 of them occur within the identified rheme and 6 of them occur outside the identified rheme (they occurred on the identified theme). Among these six gestures, were one deictic, two beat, and three metaphoric gestures.\\

%% file: implement.tex
\begin{figure}[htb]
\centering
  \includegraphics[width=3.3in]{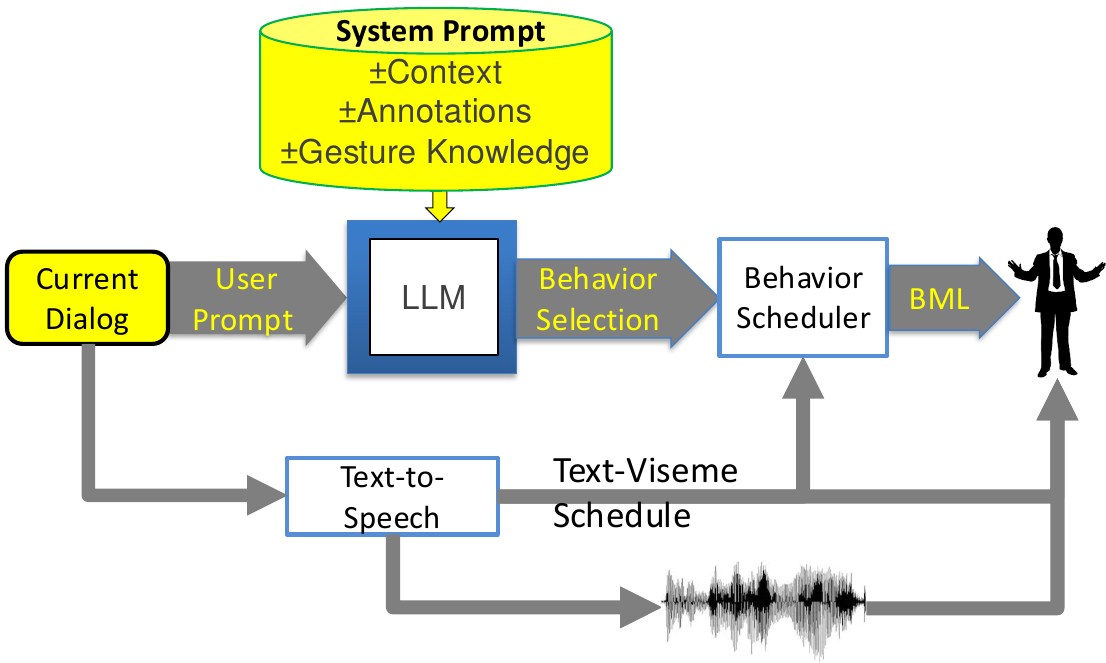}
\caption{LLM Approach to selecting gestures.\\
Based on the type of approach, the input of the LLM can contain the context, annotation (examples), or gesture knowledge (gesture description)}
\Description{gesture generation architecture}
\label{fig:arch}
\end{figure}

This section details the implementation of our proposed LLM-based nonverbal behavior generation system and explains its integration within SIMA, the Socially Intelligent Multimodal Agent. Figure~\ref{fig:arch} illustrates the architecture for selecting gestures, assuming the virtual human uses text-to-speech. Upon initiation, the LLM processes a System Prompt that includes conversational context and, potentially, examples of utterance-gesture pairings relevant to that context. The prompt may be tailored for various roles, such as a labor union representative at a panel or a presidential candidate at a rally. Furthermore, this prompting stage explores different prompting approaches to assess their impact on the creativity and accuracy of gestures proposed by the LLM.
Within our architecture, the text-to-speech and behavior scheduling processes are standard components commonly found in virtual human architectures, as described in~\cite{hartholt2022re}. The text-to-speech system generates a schedule for the behavior scheduler, which aligns nonverbal behaviors (gestures, visemes) with the corresponding audio. This behavior schedule is then sent to the animation engine in Behavior Markup Language (BML)~\cite{koppBML}, which manages co-articulation and blending of gestures. Here, we focus on how the LLM can provide specific gesture information to guide the behavior scheduler and animation engine, including labels or physical properties for real-time animation. The full implementation, and information on all prompting approaches is available on GitHub\footnote{\url{https://github.com/pariesque/SIMA}}.
\subsection{SIMA Gesture Generation}
\AddToShipoutPictureBG*{
  \AtPageLowerLeft{
    \rotatebox{90}{
      \raisebox{\dimexpr\paperheight-1cm\relax}{%
        \hspace{5cm}
        \textcolor{gray}{\Huge\textsf{This paper was submitted to AAMAS 2025 on October 16, 2024, and accepted.}}%
      }
    }
  }
}
The gesture generation component in SIMA takes speech in textual format and generates a BML file specifying the gestures that the virtual human should perform. The model used in SIMA is GPT-4, which is prompted with prompting approach 3, as specified earlier.
\subsubsection{Input Processing}
The input dialogue text is tokenized and marked with <mark name=""> tags in XML format. These <mark> tags allow the text-to-speech engine to replace markers with precise BML timings, enabling synchronization between spoken words and gestures. In the current implementation, GPT-4 directly performs the structural and semantic analyses of the dialogue text relevant to its proposed gestures. 
\subsubsection{Gesture Prompting}
Using predefined prompts, GPT-4 identifies potential gestures for corresponding parts of the utterance, specifying gestural intents (e.g., container) and their associated phrases (part of the speech where the gesture occurs). Due to how the animation system handles gestures, this step does not include the physical properties of gestures. The GPT4's output is in JSON format, which is then converted into BML. The gestural intent identified by the model is set as the gesture lexeme in the BML, and the first word in the associated phrase (identified by its word number within the complete sentence) in the output of the LLM is set as the stroke-start point in the BML. All gestures generated by GPT-4 are embedded in a single BML file, which is then passed as input to the animation engine. The animation engine then executes the provided gestures according to the provided timings (the stroke-start points).
The formatting of the behavior BML block is as follows: 
<gesture stroke-start="T3" lexeme="Container" type="METAPHORIC" emotion="neutral" />
\subsection{Animation Realization}
SIMA uses SmartBody ~\cite{thiebaux2008smartbody, shapiro2011building} as its animation engine, an open-source framework for real-time animation of conversational agents.
SmartBody converts the BML generated by the SIMA gesture generation into character animation aligned with the speech audio. SmartBody's behavior processing consists of a behavior \& and schedule manager and a motion controller engine. The behavior \& schedule manager parses the BML, extracting behavior requests and their synchronization points. These behaviors encompass gestures, speech visemes, and other nonverbal cues. This system retrieves the timing of the speech markers to synchronize the speech with the behaviors. Then, the scheduler assigns the absolute timing to the behaviors' time markers. Meanwhile, the motion controller manipulates skeletal joints, coordinating movements like gaze, gestures, and head nods to ensure fluid, synchronized animation.

The overall pipeline is automated, from the utterance to the crafting of speech audio, gesture selection by LLM, mapping to the BML specification of the animation, and realization in the SmartBody animation system.

%% file: discussion.tex
In this paper, we evaluated the use of LLMs to automate gesture selection, examining various prompting approaches that yield diverse results suitable for different applications. As to be expected, Approach 0, where the model was not prompted with any prior information, had the most flexibility in the gestures it recommended and could generate creative variations of gestures. However, this approach had the highest number of inappropriate gestures, the lowest appropriate gestures, and the lowest alignment with the speaker's gestural performance and gestural timings. One might imagine it being useful for suggesting gestures to a designer, as opposed to automating gesture selection. In other approaches, adding information to the prompts, including the explanation of gestural intents and examples from the speaker, increased the number of appropriate gestures and decreased the number of inappropriate gestures. Also, gestures tended to be more aligned with original speaker, with slight variations between approaches 1, 2 and 3.

Looking at the results provided by these different approaches, several qualitative distinctions were apparent. Note that all approaches provided, in addition to the gesture, a description of the physical properties. In general, this was very surprising, as it represented deep relationships between meaning in the utterance and the physical motion, as argued by gesture researchers~\cite{kendon2004gesture,calbris2011elements}. 

In Approach 0, which lacked gestural intents or annotation examples, the model sometimes produced shallow responses, such as self-referential deictics triggered by words like 'I,' 'we, and 'our.' 
\balance
Additionally, this approach frequently generated inappropriate iconic gestures that detracted from speech. Approach 1, which was prompted with gestural intents and descriptions, had fewer gestures than one might expect. Also sometimes the gestural intent was off while the description of the physical motion was good. Approach 2, which was prompted with annotations, but no list or description of gestural intents, would come up with novel gestural intents not in the annotations. 
Additionally, our findings on the rheme and theme analysis suggest that this linguistic construct can be used to narrow down the selection of gestures from an LLM, focusing on those most likely to match the original speaker's actions.

Looking at these results from the perspective of building a virtual agent, the animation system Figure \ref{fig:arch} requires a mapping between gesture labels (intents) and a defined set of animations. Therefore, limiting the model's choice of gestures as was done is aligned with such an animation system. Accordingly, we used approach 3 in our implementation. The generated gestures are in this case constrained to the gestural intents and annotations in our sample of the speakers' style of gesturing, which in some cases is a positive outcome in terms of personality/role consistency in behavior but in others may be viewed as limiting, especially compared to end-to-end machine learning based approaches to gesture animation. Though to be clear, what is limited here is the space of gestural types, not the language that triggers those gestures which is very general due to the use of an LLM. Given the fact that there is a many-to-many mapping between language and gesture, especially when considering an individual's gestural style, this arguably is not as limiting as it might seem.

\subsection{Open Issues}
While GPT-4 has demonstrated promising capabilities in gesture selection, its latency is too high for real-time applications in face-to-face conversations, where inference times need to be under a second, considering additional latencies in the system. To address this, we plan to explore and evaluate the fine-tuning of smaller language models to determine their suitability for integration into a virtual human framework, aiming for real-time gesture generation.

Another issue arises when there are multiple gestures in an utterance. The spatial location of gestures, and connecting one to the other becomes an important factor in ensuring the consistency of the gestures and making sure the right information is being conveyed.
For example, GPT-4 sometimes identified multiple gestures where motion of one gesture is meaningfully tied to the preceding gesture, such as a container gesture along with a gesture suggesting something being removed or placed in the container~\cite{lhommet2014metaphoric}.

This raises the issue as to the capabilities of GPT-4 to specify spatial relations between gestures in ideational units. To automatically animate, it is necessary to identify where gestures begin and end in the gestural space. 
The following explores this issue through a prompt that specifically requests spatial information.

\begin{footnotesize}
\begin{verbatim}
Prompt:
In the following utterance, please identify any Image
Schemas evoked by the utterance, the Phrase of the 
utterance that comprise the image schema and what
gesture might be used to illustrate that Phrase as well
as where that gesture Spatially Begins and 
Spatially Ends, center, left or right.
Please provide the answer in the following format:
[Type of Image Schema:]
	[Phrase : Gesture : Spatially Begins: Spatially Ends]
	[Phrase : Gesture : Spatially Begins: Spatially Ends]
	[Phrase : Gesture : Spatially Begins: Spatially Ends]
Here is the utterance:
“For example, up in New York with SEIU-1199, you know,
coming together with workers and employers putting a
little bit of contribution from each into a trust fund
to provide for care for -- whether it’s, you know,
elderly or children or whatever, to subsidize that.”

GPT4:
To analyze the utterance for image schemas and propose
gestures that illustrate the phrases, we can identify
several key elements in the text:
[Type of Image Schema: CONTAINER]
["with SEIU-1199" : Cupped hands gesture : 
   Spatially Begins:Center : Spatially Ends:Center]
["into a trust fund" : Gesture of placing objects into a 
   container (using both hands) : 
   Spatially Begins:Left : Spatially Ends:Center]
[Type of Image Schema: PATH]
["coming together" : Gesture of moving hands toward
   each other : 
   Spatially Begins:Left and Right : Spatially Ends:Center]
["putting a little bit of contribution" : Gesture mimicking
    placing or dropping something along a path : 
    Spatially Begins:Left : Spatially Ends:Center]
[Type of Image Schema: SUPPORT]
["provide for care" : Hands held flat, palms up, as if 
     supporting something : 
     Spatially Begins:Center : Spatially Ends:Center]
These gestures and their spatial dynamics help to visually
represent the conceptual ideas expressed in the text.
\end{verbatim}
\end{footnotesize}

Although above illustrations of spatially coordinating gestures in ideational units are promising, more extensive testing is required.

%% file: conclusion.tex

The assumption underlying this work is that different designers may have different intentions so their approach to automating gesture generation may differ.
In this work, we are exploring ways to automate gesture selection assuming that a designer wants to retain a degree of control over the form and the usage of the gestures.

We specifically demonstrated the use of LLMs to automate the selection of semantically rich gestures for virtual humans. As part of the effort, the approach considered ways of tailoring gesture selection to a specific individual's or role's use of gestures. We evaluated alternative prompting approaches that varied to the degree that prompting constrained gesture use. The approaches were evaluated by their semantic appropriateness (RQ1). 
In general, the semantic appropriateness was impressive regardless of whether the LLM was prompted to restrict its gesture use to those gestures common to a specific speaker.
We evaluated as well the approaches in terms of the speed of inference (RQ1), in order to determine the suitability for incremental real-time gesture selection as opposed to offline gesture design, as well the capability of the LLM to assess when to gesture (RQ2).  We additionally presented initial explorations in the ability to realize spatially consistent gestures in ideational units. Those results were very promising but need further evaluation of methods for the LLM to control the use of gestural space across gestures.
And finally, the overall approach was implemented within a virtual human architecture.

Our future steps would be to explore adapting a smaller language model to enable real-time inference while also compensating for decreases in inference appropriateness.
 

%% file: main.bbl

\begin{thebibliography}{43}


\ifx \showCODEN    \undefined \def \showCODEN     #1{\unskip}     \fi
\ifx \showDOI      \undefined \def \showDOI       #1{#1}\fi
\ifx \showISBNx    \undefined \def \showISBNx     #1{\unskip}     \fi
\ifx \showISBNxiii \undefined \def \showISBNxiii  #1{\unskip}     \fi
\ifx \showISSN     \undefined \def \showISSN      #1{\unskip}     \fi
\ifx \showLCCN     \undefined \def \showLCCN      #1{\unskip}     \fi
\ifx \shownote     \undefined \def \shownote      #1{#1}          \fi
\ifx \showarticletitle \undefined \def \showarticletitle #1{#1}   \fi
\ifx \showURL      \undefined \def \showURL       {\relax}        \fi
\providecommand\bibfield[2]{#2}
\providecommand\bibinfo[2]{#2}
\providecommand\natexlab[1]{#1}
\providecommand\showeprint[2][]{arXiv:#2}

\bibitem[\protect\citeauthoryear{Bavelas}{Bavelas}{1994}]%
        {bavelas1994gestures}
\bibfield{author}{\bibinfo{person}{Janet~Beavin Bavelas}.} \bibinfo{year}{1994}\natexlab{}.
\newblock \showarticletitle{Gestures as part of speech: Methodological implications}.
\newblock \bibinfo{journal}{\emph{Research on language and social interaction}} \bibinfo{volume}{27}, \bibinfo{number}{3} (\bibinfo{year}{1994}), \bibinfo{pages}{201--221}.
\newblock


\bibitem[\protect\citeauthoryear{Calbris}{Calbris}{2011}]%
        {calbris2011elements}
\bibfield{author}{\bibinfo{person}{Genevi{\`e}ve Calbris}.} \bibinfo{year}{2011}\natexlab{}.
\newblock \bibinfo{booktitle}{\emph{Elements of meaning in gesture}}. Vol.~\bibinfo{volume}{5}.
\newblock \bibinfo{publisher}{John Benjamins Publishing}.
\newblock


\bibitem[\protect\citeauthoryear{Cassell, Pelachaud, Badler, Steedman, Achorn, Becket, Douville, Prevost, and Stone}{Cassell et~al\mbox{.}}{1994}]%
        {cassell1994animated}
\bibfield{author}{\bibinfo{person}{Justine Cassell}, \bibinfo{person}{Catherine Pelachaud}, \bibinfo{person}{Norman Badler}, \bibinfo{person}{Mark Steedman}, \bibinfo{person}{Brett Achorn}, \bibinfo{person}{Tripp Becket}, \bibinfo{person}{Brett Douville}, \bibinfo{person}{Scott Prevost}, {and} \bibinfo{person}{Matthew Stone}.} \bibinfo{year}{1994}\natexlab{}.
\newblock \showarticletitle{Animated conversation: rule-based generation of facial expression, gesture \& spoken intonation for multiple conversational agents}. In \bibinfo{booktitle}{\emph{Proceedings of the 21st annual conference on Computer graphics and interactive techniques}}. \bibinfo{pages}{413--420}.
\newblock


\bibitem[\protect\citeauthoryear{Cassell, Stone, and Yan}{Cassell et~al\mbox{.}}{2000}]%
        {cassell2000}
\bibfield{author}{\bibinfo{person}{Justine Cassell}, \bibinfo{person}{Matthew Stone}, {and} \bibinfo{person}{Hao Yan}.} \bibinfo{year}{2000}\natexlab{}.
\newblock \showarticletitle{Coordination and context-dependence in the generation of embodied conversation}. In \bibinfo{booktitle}{\emph{Proceedings of the first international conference on {Natural} language generation - {Volume} 14}} \emph{(\bibinfo{series}{{INLG} '00})}. \bibinfo{publisher}{Association for Computational Linguistics}, \bibinfo{address}{USA}, \bibinfo{pages}{171--178}.
\newblock
\showISBNx{978-965-90296-0-0}
\urldef\tempurl%
\url{https://doi.org/10.3115/1118253.1118277}
\showDOI{\tempurl}


\bibitem[\protect\citeauthoryear{Cassell, Vilhj{\'a}lmsson, and Bickmore}{Cassell et~al\mbox{.}}{2001}]%
        {cassell2001beat}
\bibfield{author}{\bibinfo{person}{Justine Cassell}, \bibinfo{person}{Hannes~H{\"o}gni Vilhj{\'a}lmsson}, {and} \bibinfo{person}{Timothy Bickmore}.} \bibinfo{year}{2001}\natexlab{}.
\newblock \showarticletitle{Beat: the behavior expression animation toolkit}. In \bibinfo{booktitle}{\emph{Proceedings of the 28th annual conference on Computer graphics and interactive techniques}}. \bibinfo{pages}{477--486}.
\newblock


\bibitem[\protect\citeauthoryear{Cienki}{Cienki}{2005}]%
        {cienki2005image}
\bibfield{author}{\bibinfo{person}{Alan Cienki}.} \bibinfo{year}{2005}\natexlab{}.
\newblock \showarticletitle{Image schemas and gesture}.
\newblock \bibinfo{journal}{\emph{From perception to meaning: Image schemas in cognitive linguistics}}  \bibinfo{volume}{29} (\bibinfo{year}{2005}), \bibinfo{pages}{421--442}.
\newblock


\bibitem[\protect\citeauthoryear{Cienki and Koenig}{Cienki and Koenig}{1998}]%
        {cienki1998metaphoric}
\bibfield{author}{\bibinfo{person}{Alan~J Cienki} {and} \bibinfo{person}{Jean-Pierre Koenig}.} \bibinfo{year}{1998}\natexlab{}.
\newblock \showarticletitle{Metaphoric gestures and some of their relations to verbal metaphoric expressions}.
\newblock \bibinfo{journal}{\emph{Discourse and cognition: Bridging the gap}} (\bibinfo{year}{1998}), \bibinfo{pages}{189--204}.
\newblock


\bibitem[\protect\citeauthoryear{Clough and Duff}{Clough and Duff}{2020}]%
        {clough_role_2020}
\bibfield{author}{\bibinfo{person}{Sharice Clough} {and} \bibinfo{person}{Melissa~C. Duff}.} \bibinfo{year}{2020}\natexlab{}.
\newblock \showarticletitle{The {Role} of {Gesture} in {Communication} and {Cognition}: {Implications} for {Understanding} and {Treating} {Neurogenic} {Communication} {Disorders}}.
\newblock \bibinfo{journal}{\emph{Frontiers in Human Neuroscience}}  \bibinfo{volume}{14} (\bibinfo{year}{2020}).
\newblock
\showISSN{1662-5161}
\urldef\tempurl%
\url{https://www.frontiersin.org/articles/10.3389/fnhum.2020.00323}
\showURL{%
\tempurl}


\bibitem[\protect\citeauthoryear{Fares, Grimaldi, Pelachaud, and Obin}{Fares et~al\mbox{.}}{2023}]%
        {fares2023}
\bibfield{author}{\bibinfo{person}{Mireille Fares}, \bibinfo{person}{Michele Grimaldi}, \bibinfo{person}{Catherine Pelachaud}, {and} \bibinfo{person}{Nicolas Obin}.} \bibinfo{year}{2023}\natexlab{}.
\newblock \bibinfo{title}{Zero-{Shot} {Style} {Transfer} for {Gesture} {Animation} driven by {Text} and {Speech} using {Adversarial} {Disentanglement} of {Multimodal} {Style} {Encoding}}.
\newblock
\newblock
\urldef\tempurl%
\url{https://hal.science/hal-03972415}
\showURL{%
\tempurl}


\bibitem[\protect\citeauthoryear{Foley and Gentile}{Foley and Gentile}{2010}]%
        {foley_nonverbal_2010}
\bibfield{author}{\bibinfo{person}{Gretchen~N. Foley} {and} \bibinfo{person}{Julie~P. Gentile}.} \bibinfo{year}{2010}\natexlab{}.
\newblock \showarticletitle{Nonverbal {Communication} in {Psychotherapy}}.
\newblock \bibinfo{journal}{\emph{Psychiatry (Edgmont)}} \bibinfo{volume}{7}, \bibinfo{number}{6} (\bibinfo{date}{June} \bibinfo{year}{2010}), \bibinfo{pages}{38--44}.
\newblock
\showISSN{1550-5952}
\urldef\tempurl%
\url{https://www.ncbi.nlm.nih.gov/pmc/articles/PMC2898840/}
\showURL{%
\tempurl}


\bibitem[\protect\citeauthoryear{Fries}{Fries}{1984}]%
        {fries1984theme}
\bibfield{author}{\bibinfo{person}{Udo Fries}.} \bibinfo{year}{1984}\natexlab{}.
\newblock \showarticletitle{Theme and rheme revisited}. In \bibinfo{booktitle}{\emph{Modes of interpretation: Essays presented to Ernst Leisi on the occasion of his 65th birthday}}. \bibinfo{pages}{177--192}.
\newblock


\bibitem[\protect\citeauthoryear{Goldin-Meadow and Alibali}{Goldin-Meadow and Alibali}{2013}]%
        {goldin2013gesture}
\bibfield{author}{\bibinfo{person}{Susan Goldin-Meadow} {and} \bibinfo{person}{Martha~Wagner Alibali}.} \bibinfo{year}{2013}\natexlab{}.
\newblock \showarticletitle{Gesture's role in speaking, learning, and creating language}.
\newblock \bibinfo{journal}{\emph{Annual review of psychology}}  \bibinfo{volume}{64} (\bibinfo{year}{2013}), \bibinfo{pages}{257--283}.
\newblock


\bibitem[\protect\citeauthoryear{Grady}{Grady}{1997}]%
        {grady1997foundations}
\bibfield{author}{\bibinfo{person}{Joseph Grady}.} \bibinfo{year}{1997}\natexlab{}.
\newblock \showarticletitle{Foundations of meaning: Primary metaphors and primary scenes}.
\newblock  (\bibinfo{year}{1997}).
\newblock


\bibitem[\protect\citeauthoryear{Grady}{Grady}{2005}]%
        {Grady+2005+35+56}
\bibfield{author}{\bibinfo{person}{Joseph~E. Grady}.} \bibinfo{year}{2005}\natexlab{}.
\newblock \bibinfo{booktitle}{\emph{Image schemas and perception: Refining a definition}}.
\newblock \bibinfo{publisher}{De Gruyter Mouton}, \bibinfo{address}{Berlin, New York}, \bibinfo{pages}{35--56}.
\newblock
\showISBNx{9783110197532}
\urldef\tempurl%
\url{https://doi.org/doi:10.1515/9783110197532.1.35}
\showDOI{\tempurl}


\bibitem[\protect\citeauthoryear{Guellaï, Langus, and Nespor}{Guellaï et~al\mbox{.}}{2014}]%
        {guellai_prosody_2014}
\bibfield{author}{\bibinfo{person}{Bahia Guellaï}, \bibinfo{person}{Alan Langus}, {and} \bibinfo{person}{Marina Nespor}.} \bibinfo{year}{2014}\natexlab{}.
\newblock \showarticletitle{Prosody in the hands of the speaker}.
\newblock \bibinfo{journal}{\emph{Frontiers in Psychology}}  \bibinfo{volume}{5} (\bibinfo{year}{2014}).
\newblock
\showISSN{1664-1078}
\urldef\tempurl%
\url{https://www.frontiersin.org/articles/10.3389/fpsyg.2014.00700}
\showURL{%
\tempurl}


\bibitem[\protect\citeauthoryear{Hall, Goldstein, and Ingram}{Hall et~al\mbox{.}}{2016}]%
        {hall_hands_2016}
\bibfield{author}{\bibinfo{person}{Kira Hall}, \bibinfo{person}{Donna~M. Goldstein}, {and} \bibinfo{person}{Matthew~Bruce Ingram}.} \bibinfo{year}{2016}\natexlab{}.
\newblock \showarticletitle{The hands of {Donald} {Trump}: {Entertainment}, gesture, spectacle}.
\newblock \bibinfo{journal}{\emph{HAU: Journal of Ethnographic Theory}} \bibinfo{volume}{6}, \bibinfo{number}{2} (\bibinfo{date}{Sept.} \bibinfo{year}{2016}), \bibinfo{pages}{71--100}.
\newblock
\showISSN{2575-1433}
\urldef\tempurl%
\url{https://doi.org/10.14318/hau6.2.009}
\showDOI{\tempurl}
\newblock
\shownote{Publisher: The University of Chicago Press.}


\bibitem[\protect\citeauthoryear{Hartholt, Fast, Li, Kim, Leeds, and Mozgai}{Hartholt et~al\mbox{.}}{2022}]%
        {hartholt2022re}
\bibfield{author}{\bibinfo{person}{Arno Hartholt}, \bibinfo{person}{Ed Fast}, \bibinfo{person}{Zongjian Li}, \bibinfo{person}{Kevin Kim}, \bibinfo{person}{Andrew Leeds}, {and} \bibinfo{person}{Sharon Mozgai}.} \bibinfo{year}{2022}\natexlab{}.
\newblock \showarticletitle{Re-architecting the virtual human toolkit: towards an interoperable platform for embodied conversational agent research and development}. In \bibinfo{booktitle}{\emph{Proceedings of the 22nd ACM International Conference on Intelligent Virtual Agents}}. \bibinfo{pages}{1--8}.
\newblock


\bibitem[\protect\citeauthoryear{Hensel, Yongsatianchot, Torshizi, Minucci, and Marsella}{Hensel et~al\mbox{.}}{2023}]%
        {hensel2023large}
\bibfield{author}{\bibinfo{person}{Laura~Birka Hensel}, \bibinfo{person}{Nutchanon Yongsatianchot}, \bibinfo{person}{Parisa Torshizi}, \bibinfo{person}{Elena Minucci}, {and} \bibinfo{person}{Stacy Marsella}.} \bibinfo{year}{2023}\natexlab{}.
\newblock \showarticletitle{Large language models in textual analysis for gesture selection}. In \bibinfo{booktitle}{\emph{Proceedings of the 25th International Conference on Multimodal Interaction}}. \bibinfo{pages}{378--387}.
\newblock


\bibitem[\protect\citeauthoryear{Hostetter}{Hostetter}{2011}]%
        {hostetter2011}
\bibfield{author}{\bibinfo{person}{Autumn~B. Hostetter}.} \bibinfo{year}{2011}\natexlab{}.
\newblock \showarticletitle{When do gestures communicate? {A} meta-analysis.}
\newblock \bibinfo{journal}{\emph{Psychological Bulletin}} \bibinfo{volume}{137}, \bibinfo{number}{2} (\bibinfo{year}{2011}), \bibinfo{pages}{297}.
\newblock
\showISSN{1939-1455}
\urldef\tempurl%
\url{https://doi.org/10.1037/a0022128}
\showDOI{\tempurl}
\newblock
\shownote{Publisher: US: American Psychological Association.}


\bibitem[\protect\citeauthoryear{Jamalian and Tversky}{Jamalian and Tversky}{2012}]%
        {tversky2012time}
\bibfield{author}{\bibinfo{person}{Azadeh Jamalian} {and} \bibinfo{person}{Barbara Tversky}.} \bibinfo{year}{2012}\natexlab{}.
\newblock \showarticletitle{Gestures alter thinking about time}. In \bibinfo{booktitle}{\emph{Proceedings of the Annual Meeting of the Cognitive Science Society}}, Vol.~\bibinfo{volume}{34}. \bibinfo{pages}{503--508}.
\newblock


\bibitem[\protect\citeauthoryear{Kendon}{Kendon}{1997}]%
        {kendon1997gesture}
\bibfield{author}{\bibinfo{person}{Adam Kendon}.} \bibinfo{year}{1997}\natexlab{}.
\newblock \showarticletitle{Gesture}.
\newblock \bibinfo{journal}{\emph{Annual review of anthropology}} \bibinfo{volume}{26}, \bibinfo{number}{1} (\bibinfo{year}{1997}), \bibinfo{pages}{109--128}.
\newblock


\bibitem[\protect\citeauthoryear{Kendon}{Kendon}{2004}]%
        {kendon2004gesture}
\bibfield{author}{\bibinfo{person}{Adam Kendon}.} \bibinfo{year}{2004}\natexlab{}.
\newblock \bibinfo{booktitle}{\emph{Gesture: Visible action as utterance}}.
\newblock \bibinfo{publisher}{Cambridge University Press}.
\newblock


\bibitem[\protect\citeauthoryear{Kita}{Kita}{2020}]%
        {kita2020cross}
\bibfield{author}{\bibinfo{person}{Sotaro Kita}.} \bibinfo{year}{2020}\natexlab{}.
\newblock \showarticletitle{Cross-cultural variation of speech-accompanying gesture: A review}.
\newblock \bibinfo{journal}{\emph{Speech Accompanying-Gesture}} (\bibinfo{year}{2020}), \bibinfo{pages}{145--167}.
\newblock


\bibitem[\protect\citeauthoryear{Kopp, Krenn, Marsella, Marshall, Pelachaud, Pirker, Th{\'o}risson, and Vilhj{\'a}lmsson}{Kopp et~al\mbox{.}}{2006}]%
        {koppBML}
\bibfield{author}{\bibinfo{person}{Stefan Kopp}, \bibinfo{person}{Brigitte Krenn}, \bibinfo{person}{Stacy Marsella}, \bibinfo{person}{Andrew~N. Marshall}, \bibinfo{person}{Catherine Pelachaud}, \bibinfo{person}{Hannes Pirker}, \bibinfo{person}{Kristinn~R. Th{\'o}risson}, {and} \bibinfo{person}{Hannes Vilhj{\'a}lmsson}.} \bibinfo{year}{2006}\natexlab{}.
\newblock \showarticletitle{Towards a Common Framework for Multimodal Generation: The Behavior Markup Language}. In \bibinfo{booktitle}{\emph{Intelligent Virtual Agents}}, \bibfield{editor}{\bibinfo{person}{Jonathan Gratch}, \bibinfo{person}{Michael Young}, \bibinfo{person}{Ruth Aylett}, \bibinfo{person}{Daniel Ballin}, {and} \bibinfo{person}{Patrick Olivier}} (Eds.). \bibinfo{publisher}{Springer Berlin Heidelberg}, \bibinfo{address}{Berlin, Heidelberg}, \bibinfo{pages}{205--217}.
\newblock
\showISBNx{978-3-540-37594-4}


\bibitem[\protect\citeauthoryear{Lakoff and Johnson}{Lakoff and Johnson}{2008}]%
        {lakoff2008metaphors}
\bibfield{author}{\bibinfo{person}{George Lakoff} {and} \bibinfo{person}{Mark Johnson}.} \bibinfo{year}{2008}\natexlab{}.
\newblock \bibinfo{booktitle}{\emph{Metaphors we live by}}.
\newblock \bibinfo{publisher}{University of Chicago press}.
\newblock


\bibitem[\protect\citeauthoryear{Lee and Marsella}{Lee and Marsella}{2006}]%
        {lee2006nonverbal}
\bibfield{author}{\bibinfo{person}{Jina Lee} {and} \bibinfo{person}{Stacy Marsella}.} \bibinfo{year}{2006}\natexlab{}.
\newblock \showarticletitle{Nonverbal behavior generator for embodied conversational agents}. In \bibinfo{booktitle}{\emph{International Conference on Intelligent Virtual Agents}}. Springer, \bibinfo{pages}{243--255}.
\newblock


\bibitem[\protect\citeauthoryear{Lhommet and Marsella}{Lhommet and Marsella}{2014}]%
        {lhommet2014metaphoric}
\bibfield{author}{\bibinfo{person}{Margot Lhommet} {and} \bibinfo{person}{Stacy Marsella}.} \bibinfo{year}{2014}\natexlab{}.
\newblock \showarticletitle{Metaphoric gestures: towards grounded mental spaces}. In \bibinfo{booktitle}{\emph{Intelligent Virtual Agents: 14th International Conference, IVA 2014, Boston, MA, USA, August 27-29, 2014. Proceedings 14}}. Springer, \bibinfo{pages}{264--274}.
\newblock


\bibitem[\protect\citeauthoryear{Lhommet, Xu, and Marsella}{Lhommet et~al\mbox{.}}{2015}]%
        {lhommet2015cerebella}
\bibfield{author}{\bibinfo{person}{Margot Lhommet}, \bibinfo{person}{Yuyu Xu}, {and} \bibinfo{person}{Stacy Marsella}.} \bibinfo{year}{2015}\natexlab{}.
\newblock \showarticletitle{Cerebella: automatic generation of nonverbal behavior for virtual humans}. In \bibinfo{booktitle}{\emph{Proceedings of the Twenty-Ninth AAAI Conference on Artificial Intelligence}}. \bibinfo{pages}{4303--4304}.
\newblock


\bibitem[\protect\citeauthoryear{Mann and Thompson}{Mann and Thompson}{1987}]%
        {mann1987rhetorical}
\bibfield{author}{\bibinfo{person}{William~C Mann} {and} \bibinfo{person}{Sandra~A Thompson}.} \bibinfo{year}{1987}\natexlab{}.
\newblock \showarticletitle{Rhetorical structure theory: Description and construction of text structures}.
\newblock In \bibinfo{booktitle}{\emph{Natural language generation: New results in artificial intelligence, psychology and linguistics}}. \bibinfo{publisher}{Springer}, \bibinfo{pages}{85--95}.
\newblock


\bibitem[\protect\citeauthoryear{Marsella, Xu, Lhommet, Feng, Scherer, and Shapiro}{Marsella et~al\mbox{.}}{2013}]%
        {marsella2013virtual}
\bibfield{author}{\bibinfo{person}{Stacy Marsella}, \bibinfo{person}{Yuyu Xu}, \bibinfo{person}{Margaux Lhommet}, \bibinfo{person}{Andrew Feng}, \bibinfo{person}{Stefan Scherer}, {and} \bibinfo{person}{Ari Shapiro}.} \bibinfo{year}{2013}\natexlab{}.
\newblock \showarticletitle{Virtual Character Performance from Speech}. In \bibinfo{booktitle}{\emph{Proceedings of the 12th ACM SIGGRAPH/Eurographics Symposium on Computer Animation}} (Anaheim, California) \emph{(\bibinfo{series}{SCA '13})}. \bibinfo{publisher}{ACM}, \bibinfo{address}{New York, NY, USA}, \bibinfo{pages}{25--35}.
\newblock
\showISBNx{978-1-4503-2132-7}
\urldef\tempurl%
\url{https://doi.org/10.1145/2485895.2485900}
\showDOI{\tempurl}


\bibitem[\protect\citeauthoryear{McNeill}{McNeill}{1992}]%
        {mcneill1992hand}
\bibfield{author}{\bibinfo{person}{David McNeill}.} \bibinfo{year}{1992}\natexlab{}.
\newblock \bibinfo{booktitle}{\emph{Hand and mind: What gestures reveal about thought}}.
\newblock \bibinfo{publisher}{University of Chicago press}.
\newblock


\bibitem[\protect\citeauthoryear{McNeill and Duncan}{McNeill and Duncan}{2000}]%
        {mcneill2000growth}
\bibfield{author}{\bibinfo{person}{David McNeill} {and} \bibinfo{person}{Susan Duncan}.} \bibinfo{year}{2000}\natexlab{}.
\newblock \showarticletitle{Growth points in thinking-for-speaking}.
\newblock \bibinfo{journal}{\emph{Language and gesture}} \bibinfo{number}{1987} (\bibinfo{year}{2000}), \bibinfo{pages}{141--161}.
\newblock


\bibitem[\protect\citeauthoryear{Neff}{Neff}{2016}]%
        {neff2016hand}
\bibfield{author}{\bibinfo{person}{Michael Neff}.} \bibinfo{year}{2016}\natexlab{}.
\newblock \showarticletitle{Hand gesture synthesis for conversational characters}.
\newblock \bibinfo{journal}{\emph{Handbook of Human Motion}} (\bibinfo{year}{2016}), \bibinfo{pages}{1--12}.
\newblock


\bibitem[\protect\citeauthoryear{Nyatsanga, Kucherenko, Ahuja, Henter, and Neff}{Nyatsanga et~al\mbox{.}}{2023}]%
        {nyatsanga2023}
\bibfield{author}{\bibinfo{person}{Simbarashe Nyatsanga}, \bibinfo{person}{Taras Kucherenko}, \bibinfo{person}{Chaitanya Ahuja}, \bibinfo{person}{Gustav~Eje Henter}, {and} \bibinfo{person}{Michael Neff}.} \bibinfo{year}{2023}\natexlab{}.
\newblock \bibinfo{title}{A {Comprehensive} {Review} of {Data}-{Driven} {Co}-{Speech} {Gesture} {Generation}}.
\newblock
\newblock
\urldef\tempurl%
\url{https://doi.org/10.1111/cgf.14776}
\showDOI{\tempurl}
\newblock
\shownote{arXiv:2301.05339 [cs].}


\bibitem[\protect\citeauthoryear{{\"O}zer and G{\"o}ksun}{{\"O}zer and G{\"o}ksun}{2020}]%
        {ozer2020gesture}
\bibfield{author}{\bibinfo{person}{Demet {\"O}zer} {and} \bibinfo{person}{Tilbe G{\"o}ksun}.} \bibinfo{year}{2020}\natexlab{}.
\newblock \showarticletitle{Gesture use and processing: A review on individual differences in cognitive resources}.
\newblock \bibinfo{journal}{\emph{Frontiers in Psychology}}  \bibinfo{volume}{11} (\bibinfo{year}{2020}), \bibinfo{pages}{573555}.
\newblock


\bibitem[\protect\citeauthoryear{Ravenet, Pelachaud, Clavel, and Marsella}{Ravenet et~al\mbox{.}}{2018}]%
        {ravenet2018automating}
\bibfield{author}{\bibinfo{person}{Brian Ravenet}, \bibinfo{person}{Catherine Pelachaud}, \bibinfo{person}{Chlo{\'e} Clavel}, {and} \bibinfo{person}{Stacy Marsella}.} \bibinfo{year}{2018}\natexlab{}.
\newblock \showarticletitle{Automating the production of communicative gestures in embodied characters}.
\newblock \bibinfo{journal}{\emph{Frontiers in psychology}}  \bibinfo{volume}{9} (\bibinfo{year}{2018}).
\newblock


\bibitem[\protect\citeauthoryear{Saund and Marsella}{Saund and Marsella}{2021}]%
        {saund2021gesture}
\bibfield{author}{\bibinfo{person}{Carolyn Saund} {and} \bibinfo{person}{Stacy Marsella}.} \bibinfo{year}{2021}\natexlab{}.
\newblock \showarticletitle{Gesture generation}.
\newblock In \bibinfo{booktitle}{\emph{The Handbook on Socially Interactive Agents: 20 years of Research on Embodied Conversational Agents, Intelligent Virtual Agents, and Social Robotics Volume 1: Methods, Behavior, Cognition}}. \bibinfo{pages}{213--258}.
\newblock


\bibitem[\protect\citeauthoryear{Seizer}{Seizer}{2011}]%
        {seizer_uses_2011}
\bibfield{author}{\bibinfo{person}{Susan Seizer}.} \bibinfo{year}{2011}\natexlab{}.
\newblock \showarticletitle{On the {Uses} of {Obscenity} in {Live} {Stand}-{Up} {Comedy}}.
\newblock \bibinfo{journal}{\emph{Anthropological Quarterly}} \bibinfo{volume}{84}, \bibinfo{number}{1} (\bibinfo{year}{2011}), \bibinfo{pages}{209--234}.
\newblock
\showISSN{0003-5491}
\urldef\tempurl%
\url{https://www.jstor.org/stable/41237487}
\showURL{%
\tempurl}
\newblock
\shownote{Publisher: The George Washington University Institute for Ethnographic Research.}


\bibitem[\protect\citeauthoryear{Shapiro}{Shapiro}{2011}]%
        {shapiro2011building}
\bibfield{author}{\bibinfo{person}{Ari Shapiro}.} \bibinfo{year}{2011}\natexlab{}.
\newblock \showarticletitle{Building a character animation system}. In \bibinfo{booktitle}{\emph{Motion in Games: 4th International Conference, MIG 2011, Edinburgh, UK, November 13-15, 2011. Proceedings 4}}. Springer, \bibinfo{pages}{98--109}.
\newblock


\bibitem[\protect\citeauthoryear{Thiebaux, Marsella, Marshall, and Kallmann}{Thiebaux et~al\mbox{.}}{2008}]%
        {thiebaux2008smartbody}
\bibfield{author}{\bibinfo{person}{Marcus Thiebaux}, \bibinfo{person}{Stacy Marsella}, \bibinfo{person}{Andrew~N Marshall}, {and} \bibinfo{person}{Marcelo Kallmann}.} \bibinfo{year}{2008}\natexlab{}.
\newblock \showarticletitle{Smartbody: Behavior realization for embodied conversational agents}. In \bibinfo{booktitle}{\emph{Proceedings of the 7th international joint conference on Autonomous agents and multiagent systems-Volume 1}}. \bibinfo{pages}{151--158}.
\newblock


\bibitem[\protect\citeauthoryear{Tversky and Hard}{Tversky and Hard}{2009}]%
        {tversky2009embodied}
\bibfield{author}{\bibinfo{person}{Barbara Tversky} {and} \bibinfo{person}{Bridgette~Martin Hard}.} \bibinfo{year}{2009}\natexlab{}.
\newblock \showarticletitle{Embodied and disembodied cognition: Spatial perspective-taking}.
\newblock \bibinfo{journal}{\emph{Cognition}} \bibinfo{volume}{110}, \bibinfo{number}{1} (\bibinfo{year}{2009}), \bibinfo{pages}{124--129}.
\newblock


\bibitem[\protect\citeauthoryear{Zhang, Ji, Gao, and Li}{Zhang et~al\mbox{.}}{2023}]%
        {zhang2023diffmotion}
\bibfield{author}{\bibinfo{person}{Fan Zhang}, \bibinfo{person}{Naye Ji}, \bibinfo{person}{Fuxing Gao}, {and} \bibinfo{person}{Yongping Li}.} \bibinfo{year}{2023}\natexlab{}.
\newblock \showarticletitle{DiffMotion: Speech-driven gesture synthesis using denoising diffusion model}. In \bibinfo{booktitle}{\emph{International Conference on Multimedia Modeling}}. Springer, \bibinfo{pages}{231--242}.
\newblock


\bibitem[\protect\citeauthoryear{Zhang, Ji, Gao, Zhao, Wu, Jiang, Du, Ye, Zhu, Zhong, et~al\mbox{.}}{Zhang et~al\mbox{.}}{2024}]%
        {zhang2024dim}
\bibfield{author}{\bibinfo{person}{Fan Zhang}, \bibinfo{person}{Naye Ji}, \bibinfo{person}{Fuxing Gao}, \bibinfo{person}{Bozuo Zhao}, \bibinfo{person}{Jingmei Wu}, \bibinfo{person}{Yanbing Jiang}, \bibinfo{person}{Hui Du}, \bibinfo{person}{Zhenqing Ye}, \bibinfo{person}{Jiayang Zhu}, \bibinfo{person}{WeiFan Zhong}, {et~al\mbox{.}}} \bibinfo{year}{2024}\natexlab{}.
\newblock \showarticletitle{DiM-Gesture: Co-Speech Gesture Generation with Adaptive Layer Normalization Mamba-2 framework}.
\newblock \bibinfo{journal}{\emph{arXiv preprint arXiv:2408.00370}} (\bibinfo{year}{2024}).
\newblock


\end{thebibliography}
